\documentclass[pra,twocolumn,graphicx]{revtex4}
\usepackage{amssymb}
\usepackage{epsfig}
\usepackage{bm}
\usepackage{amsmath}

\begin{document}

\title{Laser-induced nonsequential double ionization at and above the
recollision-excitation-tunneling threshold}
\author{T. Shaaran$^{1},$ M. T. Nygren$^{1,2}$ and C. Figueira de Morisson
Faria$^{1}$}
\address{$^1$Department of Physics and Astronomy,
University College London, Gower Street, London WC1E 6BT, United
Kingdom\\$^2$Theoretical Physics, Blackett Laboratory, Imperial
College London, London SW7 2AZ, United Kingdom}
\date{\today}

\begin{abstract}
We perform a detailed analysis of the
recollision-excitation-tunneling (RESI) mechanism in laser-induced
nonsequential double ionization (NSDI), in which the first electron,
upon return, promotes a second electron to an excited state, from
which it subsequently tunnels, based on the strong-field
approximation. We show that the shapes of the electron momentum
distributions carry information about the bound-state with which the
first electron collides, the bound state to which the second
electron is excited, and the type of electron-electron interaction.
Furthermore, one may define a driving-field intensity threshold for
the RESI physical mechanism. At the threshold, the kinetic energy of
the first electron, upon return, is just sufficient to excite the
second electron. We compute the distributions for helium and argon
in the threshold and above-threshold intensity regime. In the latter
case, we relate our findings to existing experiments. The
electron-momentum distributions encountered are symmetric with
respect to all quadrants of the plane spanned by the momentum
components parallel to the laser-field polarization, instead of
concentrating on only the second and fourth quadrants.
\end{abstract}
\maketitle

\section{Introduction}

Electron-electron correlation in strong laser fields has raised
considerable interest for over a decade, in particular in the
context of laser-induced nonsequential double and multiple
ionization \cite{Electron correlation}. Concrete examples are the
early measurements of a \textquotedblleft knee" in the double
ionization yield as a function of the laser-field intensity, which
deviates from the predictions of sequential models in orders of
magnitude \cite{knee}, and the peaks in the electron momentum
distributions in nonsequential double ionization (NSDI), as
functions of the electron components $p_{n\parallel }$ $(n=1,2)$
parallel to the laser-field polarization \cite{Weber}. Such peaks
occur at nonvanishing parallel momenta and cannot be explained by a
sequential mechanism.

Presently, it is an established fact that NSDI occurs due to the
inelastic recollision of an electron with its parent ion
\cite{tstep}. In this recollision, the first electron gives part of
the kinetic energy it acquired from the driving field to a second
electron, which is then freed.

The simplest type of recollision which can lead to this phenomenon
is electron-impact ionization. Thereby, the first electron, upon
return, provides the second electron with enough energy so that it
is able to overcome the second ionization potential of the target in
question and reach the continuum. Both electrons leave
simultaneously and lead to distributions peaked at nonvanishing
momenta, occupying the first and third quadrant of the plane
$p_{1\parallel }$$p_{2\parallel }$ spanned by the parallel momentum
components.

Electron-impact ionization is possibly the most extensively
investigated NSDI mechanism
\cite{NSDIsfa,Carla1,Carla2,Carla3,scatteringtdse,experimentsV,Emanouil,prauzner,Denys}.
This is due to the fact that, until the past few years, it was
sufficient to describe key features observed experimentally in the
electron momentum distributions. Concrete examples are the peaks
near the non-vanishing momenta $p_{1\parallel }=p_{2\parallel }=\pm
2\sqrt{U_{p}}$, where $U_p$ is the ponderomotive energy, and the
recently reported V-shaped structure \cite{experimentsV}, which is a
signature of the long-range character of the electron-electron
interaction \cite{Carla2,Emanouil,prauzner,Denys,scatteringtdse}.
Moreover, from the theoretical viewpoint, this mechanism is
considerably easier to model, as compared to other rescattering
processes. This is particularly true in the context of
semi-analytical models, such as the strong-field approximation
\cite{NSDIsfa,Carla1,Carla2,Carla3,Kopold,Denys}.

Recent experimental results, however, reveal other rescattering
mechanisms apart from electron-impact ionization. For instance, if
the driving-field intensity is below the threshold intensity, the
kinetic energy of the recolliding electron is no longer sufficient
to make the second electron overcome the ionization potential of the
singly ionized core and reach the continuum
\cite{Eremina,threshold}. In this intensity range, the energy
transferred to the core is only enough to raise the second electron
to an excited bound state, from which it subsequently tunnels. This
process is known as recollision-excitation-tunneling ionization
(RESI). This mechanism is also important for specific targets, such
as argon \cite{Jesus}, even if intensities above the threshold are
taken, provided the driving pulse is long enough.

For this specific mechanism, the first electron leaves near a crossing of
the driving field, whilst the second electron is freed near a field maximum.
The time delay between both electrons suggests that they leave with opposite
momenta. This implies that the second and the fourth quadrant of the $%
p_{1\parallel }$$p_{2\parallel}$ plane are expected to be populated
\cite{timelag}. A population in such regions of the parallel
momentum plane has also been observed for NSDI of diatomic
molecules, especially if the molecules in question are aligned
perpendicular to the laser-field polarization
\cite{NSDIalign,Chen,Baierexcited,Prauznerexc}. Furthermore, it has
been recently shown that electron-impact ionization, alone, would
not be very useful for retrieving the molecular structure in the
experimentally relevant intensity range \cite{UCL2008}. All these
examples suggest that the RESI mechanism is becoming increasingly
important, in view of the driving field intensities and targets
employed.

This mechanism, however, is considerably less understood than
electron-impact ionization. In fact, apart from early results in
which only the RESI yield has been calculated  within a
semi-analytical framework \cite{Kopold}, most existing results are
the outcome of computations, either classical
\cite{Prauznerexc,Chen,timelag,Agapi09,otherclassical} or quantum
mechanical \cite{Baierexcited}, for which the different rescattering
mechanisms cannot be easily disentangled.

In previous work \cite{SF2009}, we have studied the RESI physical
mechanism, within the strong-field approximation. In this framework
the transition amplitude is written as a multiple integral, with a
time-dependent action and slowly-varying prefactors. This approach
has the particular advantage of providing a clear space-time picture
of the process in question, in terms of electron trajectories, while
retaining features such as quantum-interference effects
\cite{Orbits}.

We have shown that the RESI mechanism may be understood as the
combination of two processes. For the first electron, a behavior
similar to rescattered above-threshold ionization is present. The
main difference lies on the fact that part of the kinetic energy of
the first electron is given to the parent ion. This leads to a
maximal kinetic energy slightly smaller than the high-order ATI
cutoff $10U_{p}.$ The second electron, which is tunnel ionized at a
subsequent time, behaves in the same way as in direct
above-threshold ionization. Hence, its maximal kinetic energy is
given by the direct ATI cutoff, i.e., $2U_{p}.$ This allowed one to
determine kinematic constraints for the RESI mechanism, which lead
to distributions which equally occupy the four quadrants of the
$p_{1\parallel }p_{2\parallel }$ plane. For an early derivation of
somewhat different kinematic constraints for this mechanism see,
e.g., Ref.~\cite{Constraints}.

In the above-stated investigations, however, we assumed the
prefactors in the strong-field approximation amplitude to be
constant. Nonetheless, in a more realistic scenario, one should take
into account that such prefactors cause a bias in momentum space.
This may lead to electron momentum distributions concentrated on the
second and fourth quadrants of the $p_{1\parallel }p_{2\parallel }$
plane.

This work is organized as follows: In Sec. \ref{theory}, we will briefly
recall the SFA transition amplitude for the RESI mechanism, which has been
derived in \cite{SF2009}. We will start from the general expressions (Sec. %
\ref{transampl}), which will be solved by saddle-point methods. The
pertaining saddle-point equations are derived in Sec. \ref{saddle}.
Subsequently, in Sec. \ref{prefactors} we will provide the specific
prefactors employed for the hydrogenic systems to be investigated in
this work. In Sec. \ref{results}, we employ this approach to compute
electron momentum distributions for Helium and Argon. For the latter
species, an explicit comparison with the results in
Ref.~\cite{Eremina} is performed. Finally, in Sec. \ref{conclusions}
we state the main conclusions of this paper.

\section{Transition amplitude}

\label{theory}

\subsection{General expressions}

\label{transampl}

The transition amplitude describing the
recollision-excitation-tunneling ionization (RESI) mechanism, within
the SFA, can be written as (for details on the derivation see
\cite{SF2009}).
\begin{eqnarray}
M(\mathbf{p}_{1},\mathbf{p}_{2}) &=&\int_{-\infty }^{\infty }\hspace*{-0.2cm}%
dt\hspace*{-0.1cm}\int_{-\infty }^{t}\hspace*{-0.3cm}dt^{\prime }\hspace*{%
-0.1cm}\int_{-\infty }^{t^{\prime }}\hspace*{-0.3cm}dt^{^{\prime \prime }}%
\hspace*{-0.1cm}\int d^{3}k  \notag \\
&&V_{\mathbf{p}_{2}e}V_{\mathbf{p}_{1}e,\mathbf{k}g}V_{\mathbf{k}g}e^{iS(%
\mathbf{p}_{1},\mathbf{p}_{2},\mathbf{k},t,t^{\prime },t^{\prime \prime })},
\label{Mp}
\end{eqnarray}
with the action
\begin{eqnarray}
S(\mathbf{p}_{1},\mathbf{p}_{2},\mathbf{k},t,t^{\prime },t^{^{\prime \prime
}}) &\hspace*{-0.12cm}=\hspace*{-0.12cm}&-\int_{t}^{\infty }\hspace{-0.1cm}\frac{[\mathbf{p}_{2}+\mathbf{A}%
(\tau )]^{2}}{2}d\tau   \notag \\
&&-\int_{t^{^{\prime }}}^{\infty }\hspace{-0.1cm}\frac{[\mathbf{p}_{1}+%
\mathbf{A}(\tau )]^{2}}{2}d\tau   \notag \\
&&-\int_{t^{^{^{\prime \prime }}}}^{t^{^{\prime }}}\hspace{-0.1cm}\frac{[%
\mathbf{k}+\mathbf{A}(\tau )]^{2}}{2}d\tau   \notag \\
&&+E_{1g}t^{^{\prime \prime }}+E_{2g}t^{^{\prime }}+E_{2e}(t-t^{^{\prime }})
\label{singlecS}
\end{eqnarray}
and the prefactors%
\begin{eqnarray}
V_{\mathbf{k}g} &=&\left\langle \mathbf{\tilde{k}}(t^{\prime \prime
})\right\vert V\left\vert \psi _{g}^{(1)}\right\rangle =\frac{1}{(2\pi
)^{3/2}}  \notag \\
&&\times \int d^{3}r_{1}V(r_{1})\exp [-i\mathbf{\tilde{k}}(t^{\prime
\prime })\cdot \mathbf{r}_{1}]\psi _{g}^{(1)}(\mathbf{r}_{1}),
\label{Vkg}
\end{eqnarray}%
\begin{eqnarray}
V_{\mathbf{p}_{1}e\mathbf{,k}g} &=&\left\langle \mathbf{\tilde{p}}_{1}\left(
t^{\prime }\right) ,\psi _{e}^{(2)}\right\vert V_{12}\left\vert \mathbf{%
\tilde{k}}(t^{\prime }),\psi _{g}^{(2)}\right\rangle =\frac{1}{(2\pi )^{3}}
\notag \\
&&\times \int \int d^{3}r_{2}d^{3}r_{1}\exp [-i(\mathbf{p}_{1}-\mathbf{k}%
)\cdot \mathbf{r}_{1}]  \notag \\
&&\times V_{12}(\mathbf{r}_{1,}\mathbf{r}_{2})[\psi _{e}^{(2)}(\mathbf{r}%
_{2})]^{\ast }\psi _{g}^{(2)}(\mathbf{r}_{2})  \label{Vp1e,kg}
\end{eqnarray}
and
\begin{eqnarray}
V_{\mathbf{p}_{2}e} &=&\left\langle \mathbf{\tilde{p}}_{2}\left( t\right)
\right\vert V_{\mathrm{ion}}\left\vert \psi _{e}^{(2)}\right\rangle =\frac{1%
}{(2\pi )^{3/2}}  \notag \\
&&\times \int d^{3}r_{2}V_{\mathrm{ion}}(r_{2})\exp [-i\mathbf{\tilde{p}}%
_{2}(t)\cdot \mathbf{r}_{2}]\psi _{g}^{(2)}(\mathbf{r}_{2}).
\label{Vp2e}
\end{eqnarray}

Eq. (\ref{Mp}) describes the physical process in which an electron,
initially in a bound state $|\psi _{g}^{(1)}>,$ is released by tunneling
ionization at a time $t^{\prime \prime }$ into a Volkov state $|\mathbf{%
\tilde{k}}(t^{\prime })>$. Subsequently, this electron propagates in the
continuum from $t^{\prime \prime }$ to a later time $t^{\prime }.$ At this
time, it is driven back by the field and rescatters with its parent ion. In
this collision, it excites the second electron, which is bound at $|\psi
_{g}^{(2)}>,$ to the state $|\psi _{e}^{(2)}>$, through the interaction $%
V_{12}.$ Finally, the second electron, which is in a bound excited state $%
|\psi _{e}^{(2)}>$, is released by tunneling ionization at a time $t$ into a
Volkov state $|\mathbf{\tilde{p}}_{2}\left( t\right) >$. The final electron
momenta are described by $\mathbf{p}_{n}(n=1,2).$ In the above-stated
equations, $E_{ng}$ $(n=1,2)$ give the ionization potentials of the ground
state, $E_{ne}$ $(n=1,2)$ give the ionization potentials of the excited
state and $V(r_{1})$ \ and $V_{\mathrm{ion}}(r_{2})$ correspond to the
atomic binding potential of the system as seen by the first and second
electron, respectively.

The form factors (\ref{Vkg}) and (\ref{Vp2e}) contain all the information
about the binding potential of the first and second electron, respectively.
The form factor (\ref{Vp1e,kg}) contains all the information about the
interaction of the first electron with the singly ionized atom, which is an
inelastic interaction. If the electron interaction is only dependent on the
difference between both electron coordinates, i.e., if $V_{12}(\mathbf{r}%
_{1,}\mathbf{r}_{2})=V_{12}(\mathbf{r}_{1}-\mathbf{r}_{2}),$ Eq. (\ref%
{Vp1e,kg}) may be rewritten as
\begin{eqnarray}
V_{\mathbf{p}_{1}e\mathbf{,k}g} &=&\frac{V_{12}(\mathbf{p}_{1}-\mathbf{k})}{%
(2\pi )^{3/2}}  \notag \\
&&\times \int d^{3}r_{2}e^{-i(\mathbf{p}_{1}-\mathbf{k})\cdot \mathbf{r}%
_{2}}[\psi _{e}^{(2)}(\mathbf{r}_{2})]^{\ast }\psi _{g}^{(2)}(\mathbf{r}%
_{2}),  \label{resc1st}
\end{eqnarray}%
with
\begin{equation}
V_{12}(\mathbf{p}_{1}-\mathbf{k})=\frac{1}{(2\pi )^{3/2}}\int d^{3}r\exp [-i(%
\mathbf{p}_{1}-\mathbf{k})\cdot \mathbf{r}]V_{12}(\mathbf{r})
\end{equation}%
and $\mathbf{r=r}_{1}-\mathbf{r}_{2}.$

Clearly, $V_{\mathbf{k}g}$ and $V_{\mathbf{p}_{2}e}$\ are gauge dependent.
In fact, in the length gauge $\mathbf{\tilde{p}}_{n}\left( \tau \right) =%
\mathbf{p}_{n}+\mathbf{A}(\tau )$ and $\mathbf{\tilde{k}}(\tau )=\mathbf{k}+%
\mathbf{A}(\tau )(\tau =t^{\prime },t^{\prime \prime }),$ while in the
velocity gauge $\mathbf{\tilde{p}}_{n}\left( \tau \right) =\mathbf{p}_{n}$
and $\mathbf{\tilde{k}}(\tau )=\mathbf{k}.$ This is a direct consequence of
the fact that the gauge transformation $\chi _{l\rightarrow v}=\exp [-i%
\mathbf{A}(\tau )\cdot \mathbf{r}]$ from the length to the velocity gauge
causes a translation $\mathbf{p}\rightarrow \mathbf{p}-\mathbf{A}(\tau )$ in
momentum space. Due to the fact that these shifts cancel out in Eq. (\ref%
{resc1st}), $V_{\mathbf{p}_{1}e\mathbf{,k}g}$ remains the same in the length
and velocity gauges.

\subsection{Saddle-point analysis}

\label{saddle}

The multiple integral in Eq.~(\ref{Mp}) will be solved using saddle-point
methods (for details see Ref. \cite{uniformATI}). For that purpose, we must find the coordinates $(t_{s},t_{s}^{%
\prime },t_{s}^{\prime \prime },\mathbf{k}_{s})$ for which $S(\mathbf{p}_{1},%
\mathbf{p}_{2},\mathbf{k},t,t^{\prime },t^{\prime \prime })$ is stationary,
i.e., for which the conditions $\partial _{t}S(\mathbf{p}_{1},\mathbf{p}_{2},%
\mathbf{k},t,t^{\prime },t^{\prime \prime })=\partial _{t^{\prime }}S(%
\mathbf{p}_{1},\mathbf{p}_{2},\mathbf{k},t,t^{\prime },t^{\prime \prime
})=\partial _{t^{^{\prime \prime }}}S(\mathbf{p}_{1},\mathbf{p}_{2},\mathbf{k%
},t,t^{\prime },t^{\prime \prime })=0$ and $\partial _{\mathbf{k}}S(\mathbf{p%
}_{1},\mathbf{p}_{2},\mathbf{k},t,t^{\prime },t^{\prime \prime })=\mathbf{0}$
are satisfied. This leads to the equations

\begin{equation}
\left[ \mathbf{k}+\mathbf{A}(t^{\prime \prime })\right] ^{2}=-2E_{1g},
\label{saddle1}
\end{equation}%
\begin{equation}
\mathbf{k=}-\frac{1}{t^{\prime }-t^{\prime \prime }}\int_{t^{\prime \prime
}}^{t^{\prime }}d\tau \mathbf{A}(\tau )  \label{saddle2}
\end{equation}%
\begin{equation}
\lbrack \mathbf{p}_{1}+\mathbf{A}(t^{\prime })]^{2}=\left[ \mathbf{k}+%
\mathbf{A}(t^{\prime })\right] ^{2}-2(E_{2g}-E_{2e}).  \label{saddle3}
\end{equation}

\bigskip and%
\begin{equation}
\lbrack \mathbf{p}_{2}+\mathbf{A}(t)]^{2}=\mathbf{-}2E_{2e}  \label{saddle4}
\end{equation}

\bigskip

Eq. (\ref{saddle1}) gives the conservation of energy at the time $\
t^{\prime \prime },$ which, physically, corresponds to tunneling of the
first electron. Since tunneling has no classical counterpart, this equation
possesses no real solution. Eq. (\ref{saddle2}) constrains the intermediate
momentum $\mathbf{k}$ of the first electron, so that it can return to the
parent ion. Eq. (\ref{saddle3}) expresses the fact that the first electron
returns to its parent ion at a time $t^{\prime }$ and rescatters
inelastically with it, giving part of its kinetic energy $E_{ret}(t^{\prime
})=\left[ \mathbf{k}+\mathbf{A}(t^{\prime })\right] ^{2}/2$ to the core.
Under this interaction the second electron is excited from a state with
energy $E_{2g}$ to a state with energy $E_{2e}$. The first electron leaves
immediately and reaches the detector with momentum $\mathbf{p}_{1}$.
Finally, Eq.~(\ref{saddle4}) describes the fact that the second electron
tunnels at a later time $t$ from an excited state of energy $E_{2e}$.

\subsubsection{Momentum constraints}

The saddle-point equations (\ref{saddle3}) and (\ref{saddle4}) provide
useful information on the momentum-space regions populated by the RESI
mechanism, and on the shapes of the electron-momentum distributions. For
instance, Eq. (\ref{saddle4}) is identical to that describing tunnel
ionization for direct above-threshold ionization. This implies that the
maximal kinetic energy of the second electron at the detector, if the field
can be approximated by a monochromatic wave, is roughly given by $2U_{p}.$

Furthermore, the electron is leaving the excited state with largest
probability when the electric field $E(t)=-dA(t)/dt$ is maximum. If the time
dependence of the laser field is such that $A(t)$ vanishes when $E(t)$ is at
its peak (for instance, monochromatic fields), then
\begin{equation}
-2\sqrt{U_{p}}\leq p_{2}\leq 2\sqrt{U_{p}}.
\end{equation}
If, to first approximation, we neglect the momentum components perpendicular
to the laser-field polarization, one can see that the momentum of the second
electron, in the parallel momentum plane, is expected to be centered around
vanishing momentum $p_{2\parallel }$ and be limited by the bounds $%
p_{2\parallel }=\pm 2\sqrt{U_{p}}.$

One should note that, due to the fact that Eq. (\ref{saddle4}) describes a
tunneling process, there is no classically allowed region for the second
electron. A non-vanishing perpendicular momentum, effectively, will lead to
an increase in the potential barrier and a suppression in the yield. In
fact, Eq. (\ref{saddle4}) can also be written as
\begin{equation}
\lbrack p_{2\parallel }+A(t)]^{2}=\mathbf{-}2\tilde{E}_{2e},
\end{equation}%
where $\tilde{E}_{2e}=E_{2e}-\mathbf{p}_{2\perp }^{2}$ is an effective
ionization potential.

The saddle-point equation (\ref{saddle3}), on the other hand, yields
information about the momentum of the second electron. According to this
equation,
\begin{equation}
-A(t)-\sqrt{2E_{\mathrm{diff}}}\leq p_{1\parallel }\leq -A(t)+\sqrt{2E_{%
\mathrm{diff}}},
\end{equation}%
where $E_{\mathrm{diff}}=E_{\mathrm{kin}}(t^{\prime },t^{\prime
\prime
})-(E_{2g}-E_{2e})-\mathbf{p}_{1\perp }^{2}/2$ and $E_{\mathrm{kin}%
}(t^{\prime },t^{\prime \prime })$ denotes the kinetic energy of the first
electron upon return. For a monochromatic field, the electron returns most
probably near a crossing of the laser field, one may use the approximation $%
A(t)\simeq 2\sqrt{U_{p}}$ in the above-stated equation. In this case, we
also know that the kinetic energy $E_{\mathrm{kin}}(t^{\prime },t^{\prime
\prime })\leq 3.17U_{p}.$ Hence, $E_{\mathrm{diff}}^{(\max )}\leq
3.17U_{p}-(E_{2g}-E_{2e})-\mathbf{p}_{1\perp }^{2}/2$ and
\begin{equation}
-2\sqrt{U_{p}}-\sqrt{2E_{\mathrm{diff}}^{(\max )}}\leq p_{1\parallel }\leq -2%
\sqrt{U_{p}}+\sqrt{2E_{\mathrm{diff}}^{(\max )}}.  \label{constraintp1}
\end{equation}%
Eq. (\ref{constraintp1}) allows one to delimit a region in momentum space
for $p_{1\parallel }$ centered around $-2\sqrt{U_{p}}$ and bounded by $2E_{%
\mathrm{diff}}^{(\max )}$. In contrast to the previous case, there may be a
classically allowed region for the momentum of the first electron if the
parameters inside the square root are positive, i.e., if $3.17U_{p}\geq
(E_{2g}-E_{2e})+\mathbf{p}_{1\perp }^{2}/2.$ For increasing perpendicular
momentum and/or bound-state energy difference, this region will become more
and more localized around $-2\sqrt{U_{p}}$ until it collapses. Therefore, it
is also possible to distinguish a threshold and an above-threshold behavior
in the context of recollision-excitation-tunneling. One should note,
however, that intensities below the recollision-excitation threshold $%
(E_{2g}-E_{2e})=3.17U_{p}$ do not make physically sense, as the energy of
the returning electron would not be sufficient to promote the bound electron
to an excited state. If $(E_{2g}-E_{2e})\ll 3.17U_{p}$ the well-known cutoff
of $10U_{p}$ for rescattered above-threshold ionization is recovered
\footnote{%
Different bounds have been provided in our previous publication \cite{SF2009}%
. Such bounds have been derived based on the above-threshold ionization
kinetic energy values. Hence, they are only applicable when $%
E_{kin}(t^{\prime },t^{\prime \prime })\gg E_{2g}-E_{2e},$ while the present
bounds are valid throughout.}.

In view of the above-mentioned constraints, the expected maxima of the
electron momentum distribution are located at the most probable momenta $%
(p_{1||},p_{2||})=(\pm 2\sqrt{U_{p}},0)$, and, after symmetrizing with
respect to the exchange $\mathbf{p}_{1}\leftrightarrow \mathbf{p}_{2}$, at $%
(p_{1||},p_{2||})=(0,\pm 2\sqrt{U_{p}})$. This implies that, if the field
can be approximately described by a monochromatic wave, the outcome of our
model should be distributions in the $p_{1\parallel }p_{2\parallel }$ plane,
which are symmetric upon $\mathbf{p}_{n}\rightarrow -\mathbf{p}_{n}$, $n=1,2$
and upon $\mathbf{p}_{1}\leftrightarrow \mathbf{p}_{2}$,\ and which equally
occupy the four quadrants of the parallel momentum plane.

\subsubsection{Bound-state singularity}

Finally, due to the saddle-point equations (\ref{saddle1}) and (\ref{saddle4}%
), for exponentially decaying bound states the prefactors (\ref{Vkg}) and (%
\ref{Vp2e}) exhibit singularities in the length-gauge formulation of the
strong-field approximation. This is due to the fact that these prefactors
will be inversely proportional to $\left( \left[ \mathbf{k}+\mathbf{A}%
(t^{\prime \prime })\right] ^{2}+2E_{1g}\right) ^{n}$ and $\left( [\mathbf{p}%
_{2}+\mathbf{A}(t)]^{2}+2E_{2e}\right) ^{m},$ where $n,m$ are integers. For
the problem addressed in this specific work, however, only the prefactor $V_{%
\mathbf{p}_{2}e}$ will influence the shape of the electron momentum
distributions. The prefactor $V_{\mathbf{k}g}$ will affect the electron
momentum distributions only quantitatively. Hence, to first approximation,
one can consider Eq.~(\ref{Vkg}) as constant. A similar problem for the
electron-impact ionization mechanism in NDSI has been discussed in detail in
\cite{Carla2}.

To overcome the singularity in $V_{\mathbf{p}_{2}e},$ one needs to embed
this prefactor into the action, which now reads
\begin{equation}
\tilde{S}(\mathbf{p}_{1},\mathbf{p}_{2},\mathbf{k},t,t^{\prime },t^{\prime
\prime })=S(\mathbf{p}_{1},\mathbf{p}_{2},\mathbf{k},t,t^{\prime },t^{\prime
\prime })-i\ln V_{\mathbf{p}_{2}e}.
\end{equation}%
This will lead to modifications in the saddle-point equation $\partial _{t}%
\tilde{S}(\mathbf{p}_{1},\mathbf{p}_{2},\mathbf{k},t,t^{\prime },t^{\prime
\prime }),$ which is now given by
\begin{equation}
\lbrack \mathbf{p}_{2}+\mathbf{A}(t)]^{2}=\mathbf{-}2E_{2e}+i\partial
_{t}\ln V_{\mathbf{p}_{2}e}.  \label{saddlemod}
\end{equation}%
The main consequence of such a modification is that the drift velocity of
the second electron is no longer pure imaginary. This will lead to a
splitting in the ionization time $t$ for each orbit, as compared to the
non-modified case. Depending on the velocity in question, the barrier the
electron must tunnel through in order to reach the continuum will either
widen or narrow. This means that, with regard to the non-modified action, $%
\mathrm{Im}[t]$ will either increase or decrease.%


\subsection{Prefactors}

\label{prefactors}

In this work, we are particularly interested in exponentially decaying,
hydrogenic bound states. This means that, in general, the bound-state
wavefunction reads
\begin{equation}
\psi ^{(\alpha )}(\mathbf{r}_{\alpha })=R_{nl}(r_{n})Y_{l}^{m}(\theta
_{\alpha },\varphi _{\alpha }),  \label{hydrogenic}
\end{equation}%
where $n,l$ and $m$ denote the principal, orbital and magnetic
quantum numbers, the index $\alpha $ refers to the electron in
question, and the angular coordinates are given by $\theta _{\alpha
}$ and $\varphi _{\alpha }$. In this case, the binding potentials
$V(r_{1})$ and $V_{\mathrm{ion}}(r_{2})$ will be given by
\begin{equation}
V_{\alpha }(r_{\alpha })=-\frac{Z_{\mathrm{eff}}}{r_{\alpha }},
\end{equation}%
where $V_{\alpha }$ yields either $V$ or $V_{\mathrm{ion}},$ and $Z_{\mathrm{%
eff}}$ corresponds to the effective electronic charge. The general
expressions for the prefactors in this work are provided in the appendix.

Below, we state the specific prefactors to be employed in Sec. \ref{results}%
, for Helium and Argon. In the former case, upon collision, the second
electron may be excited from the $1s$ state to either the $2s$ or the $2p$
state, while in the latter species it may undergo a transition from the $3p$
state to the $4s$ or the $4p$ state. One should note that the prefactor $V_{%
\mathbf{p}_{2}e}$ is gauge dependent. In the length gauge, $\tilde{\mathbf{p}%
}_{2}(t)=\mathbf{p}_{2}+\mathbf{A}(t)$, and $\tilde{p}_{2}(t)_{\parallel
}=p_{2\parallel }+A(t)$ while, in the velocity gauge, $\tilde{\mathbf{p}}%
_{2}(t)=\mathbf{p}_{2}$ and $\tilde{p}_{2}(t)_{\parallel }=p_{2\parallel }$.
The prefactor $V_{\mathbf{p}_{1}e\mathbf{,k}g},$ on the other hand, is gauge
invariant.

\subsubsection{Excitation $1s\rightarrow 2s$}

Let us first consider the simplest case, in which the second electron is
excited to $2s.$ This gives the prefactors
\begin{equation}
V_{\mathbf{p}_{2}e}^{(2s)}\sim \frac{\left[ \tilde{p}_{2}(t)\right]
^{2}-2E_{2e}}{[\left[ \tilde{p}_{2}(t)\right] ^{2}+2E_{2e}]^{2}}
\end{equation}%
and
\begin{equation}
V_{\mathbf{p}_{1}e\mathbf{,k}g}^{(1s\rightarrow 2s)}\sim V_{12}(\mathbf{p}%
_{1}-\mathbf{k})\frac{\eta _{1}(\kappa ^{2},E_{2g},E_{2e})}{[\kappa
^{2}+\zeta ^{2}(E_{2g},E_{2e})]^{3}},
\end{equation}%
with
\begin{eqnarray}
\eta _{1}(\kappa ^{2},E_{2g},E_{2e}) &=&\kappa ^{2}(\sqrt{2E_{2g}}+2\sqrt{%
2E_{2e}})+\left( 2E_{2g}\right) ^{3/2}  \notag \\
&&-2(2E_{2e})^{3/2}-6E_{2e}\sqrt{2E_{2g}}.
\end{eqnarray}%
and
\begin{equation}
\zeta (E_{2g},E_{2e})=\sqrt{2E_{2e}}+\sqrt{2E_{2g}}.  \label{zeta}
\end{equation}

The above-stated equations can also be written in terms of the momentum
components parallel and perpendicular to the laser field polarization,
denoted by $p_{\alpha ||}$ and $p_{\alpha \perp },$ $(\alpha =1,2$),
respectively. In this case,

\begin{equation}
V_{\mathbf{p}_{2}e}^{(2s)}\sim \frac{\left[ \tilde{p}_{2}(t)_{\parallel }%
\right] ^{2}+\mathbf{p}_{2\perp }^{2}-2E_{2e}}{[\left[ \tilde{p}%
_{2}(t)_{\parallel }\right] ^{2}+\mathbf{p}_{2\perp }^{2}+2E_{2e}]^{2}}
\label{Vp2e2s}
\end{equation}%
\begin{equation}
V_{\mathbf{p}_{1}e\mathbf{,k}g}^{(1s\rightarrow 2s)}\sim V_{12}(\mathbf{p}%
_{1}-\mathbf{k})\frac{\eta _{1}\left[ \left( k-p_{1\parallel }\right) ^{2}+%
\mathbf{p}_{1\perp }^{2},E_{2g},E_{2e}\right] }{[\left( k-p_{1\parallel
}\right) ^{2}+\mathbf{p}_{1\perp }^{2}+\zeta ^{2}(E_{2g},E_{2e})]^{3}},
\end{equation}

\subsubsection{Excitation $1s\rightarrow 2p$}

If, on the other hand, the second electron is excited to $2p$, one must
consider three degenerate states, corresponding to the magnetic quantum
numbers $m=\pm 1,0$. 
This yields
\begin{equation}
V_{\mathbf{p}_{2}e}^{(2p)}\sim \frac{\sqrt{\left[ \tilde{p}_{2}(t)\right]
^{2}}}{\left( 2E_{2e}+\left[ \tilde{p}_{2}(t)\right] ^{2}\right) ^{2}}\left[
Y_{1}^{m}(\theta _{\tilde{p}_{2}},\varphi _{\tilde{p}_{2}})\right] ^{\ast }
\end{equation}%
and%
\begin{equation}
V_{\mathbf{p}_{1}e\mathbf{,k}g}^{(1s\rightarrow 2p)}\sim V_{12}(\mathbf{p}%
_{1}-\mathbf{k})\eta _{2}(\kappa ^{2},E_{2g},E_{2e})\left[ Y_{1}^{m}(\theta
_{\kappa },\varphi _{\kappa })\right] ^{\ast },
\end{equation}%
with
\begin{equation}
\eta _{2}(\kappa ^{2},E_{2g},E_{2e})=\frac{\zeta (E_{2g},E_{2e})\sqrt{\kappa
^{2}}}{\left( \zeta ^{2}(E_{2g},E_{2e})+\kappa ^{2}\right) ^{3}}.
\end{equation}%
Since the electron may be excited to any of the $2p$ states, we will
consider the coherent superposition
\begin{equation}
\left\vert \psi _{2p}^{(2)}\right\rangle =\frac{1}{\sqrt{3}}\left(
\left\vert \psi _{2p_{x}}^{(2)}\right\rangle +\left\vert \psi
_{2p_{y}}^{(2)}\right\rangle +\left\vert \psi _{2p_{z}}^{(2)}\right\rangle
\right) ,
\end{equation}%
where $\left\langle \mathbf{r}_{2}\right. \left\vert \psi
_{2p_{j}}^{(2)}\right\rangle =\psi _{2p_{j}}^{(2)}(\mathbf{r}_{2}),$ with $%
j=x,y,z.$ This implies that%
\begin{equation}
V_{\mathbf{p}_{2}e}^{(2p)}\sim \frac{\sqrt{\left[ \tilde{p}_{2}(t)\right]
^{2}}}{\left( 2E_{2e}+\left[ \tilde{p}_{2}(t)\right] ^{2}\right) ^{2}}\beta (%
\mathbf{\tilde{p}}_{2}(t))  \label{Vp2e1s2p}
\end{equation}%
and
\begin{equation}
V_{\mathbf{p}_{1}e\mathbf{,k}g}^{(1s\rightarrow 2p)}\sim V_{12}(\mathbf{p}%
_{1}-\mathbf{k})\eta _{2}(\kappa ^{2},E_{2g},E_{2e})\beta (\mathbf{\kappa }),
\label{Vp1k1s2p}
\end{equation}%
where the angular dependency is given by
\begin{equation}
\beta (\mathbf{q})=(\sin \theta _{q}\cos \varphi _{q}+\sin \theta _{q}\sin
\varphi _{q}+\cos \theta _{q}).
\end{equation}%
Thereby, we employed the usual relations between spherical polar coordinates
and the spherical harmonics.


One may write the above-stated expressions in terms of the electron momentum
components parallel and perpendicular to the laser-field polarization. In
this case, Eq. (\ref{Vp2e1s2p}) reads%
\begin{equation}
V_{\mathbf{p}_{2}e}^{(2p)}\sim \frac{\sqrt{\left[ \tilde{p}%
_{2}(t)_{\parallel }\right] ^{2}+\mathbf{p}_{2\perp }^{2}}}{\left( 2E_{2e}+%
\left[ \tilde{p}_{2}(t)_{\parallel }\right] ^{2}+\mathbf{p}_{2\perp
}^{2}\right) ^{2}}\beta (\mathbf{\tilde{p}}_{2}(t)).
\end{equation}%
In $\beta (\mathbf{\tilde{p}}_{2}(t)),$ the angles $\theta _{\tilde{p}_{2}}$
and $\varphi _{\tilde{p}_{2}}$ are given by
\begin{equation}
\theta _{\tilde{p}_{2}}=\arccos \left[ \tilde{p}_{2}(t)_{\parallel }/\sqrt{%
\left[ \tilde{p}_{2}(t)_{\parallel }\right] ^{2}+\mathbf{p}_{2\perp }^{2}}%
\right]
\end{equation}%
and $\varphi _{\tilde{p}_{2}}=\arccos [\tilde{p}_{2}(t)_{x}/\tilde{p}%
_{2}(t)_{\perp }],$ respectively. In Eq. (\ref{Vp1k1s2p}), $\kappa
^{2}=(k-p_{1\parallel })^{2}+\mathbf{p}_{1\perp }^{2}$ and the angles $%
\theta _{\mathbf{\kappa }}$ and $\varphi _{\mathbf{\kappa }}$ read $\ $%
\begin{equation}
\theta _{\mathbf{\kappa }}=\arccos \left[ \left( k-p_{1\parallel }\right) /%
\sqrt{(k-p_{1\parallel })^{2}+\mathbf{p}_{1\perp }^{2}}\right]
\end{equation}%
and $\varphi _{\mathbf{\kappa }}=\arccos [p_{1x}/p_{1\perp }],$
respectively. This angular dependence will be washed out when the transverse
momentum components are integrated over (see Sec. \ref{results}).

\subsubsection{Excitation $3p\rightarrow 4s$ and $3p\rightarrow 4p$}

Finally, we will assume that the second electron, initially in $3p,$ will be
excited either to the $4s$ or to the $4p$ state. Similarly to the procedure
adopted in the previous section, we will consider a coherent superposition
of the $3p_{x},$ $3p_{y}$ and $3p_{z}$ states for the initial state of the
electron, i.e.,
\begin{equation}
\left\vert \psi _{3p}^{(2)}\right\rangle =\frac{1}{\sqrt{3}}\left(
\left\vert \psi _{3p_{x}}^{(2)}\right\rangle +\left\vert \psi
_{3p_{y}}^{(2)}\right\rangle +\left\vert \psi _{3p_{z}}^{(2)}\right\rangle
\right) .
\end{equation}%
If the electron is excited to the $4s$ state, the excitation prefactor $V_{%
\mathbf{p}_{1}e\mathbf{,k}g}^{(3p\rightarrow 4s)}$ will exhibit an angular
dependence given by $\beta (\mathbf{\kappa }),$ and the tunneling prefactor $%
V_{\mathbf{p}_{2}e}^{(4s)}$ will not depend on the angular variables. Both
prefactors also have a radial dependence on $\tilde{p}_{2}(t)$ or $\kappa
\mathbf{.}$ If, however, the electron is excited to the $4p$ state, one must
take the final state as
\begin{equation}
\left\vert \psi _{4p}^{(2)}\right\rangle =\frac{1}{\sqrt{3}}\left(
\left\vert \psi _{4p_{x}}^{(2)}\right\rangle +\left\vert \psi
_{4p_{y}}^{(2)}\right\rangle +\left\vert \psi _{4p_{z}}^{(2)}\right\rangle
\right) ,
\end{equation}%
i.e., as a coherent superposition of $4p_{x},$ $4p_{y}$ and $4p_{z}.$ In
this case, the angular dependence of $V_{\mathbf{p}_{2}e}$ will be embedded
in $\beta (\mathbf{\tilde{p}}_{2}(t)).$ The angular dependence of $V_{%
\mathbf{p}_{1}e\mathbf{,k}g}^{(3p\rightarrow 4p)}$ will be more complex and
will involve the sum of the orbital angular momenta of the two electronic
bound states involved. Due to the higher quantum numbers involved, the
prefactors are messier than those in the previous sections and will not be
written down explicitly. They can, however, be obtained from the general
expressions in the appendix.

\section{Electron momentum distributions}

\label{results}

In this section, we will compute electron momentum distributions, as
functions of the momentum components $(p_{1\parallel },p_{2\parallel })$
parallel to the laser-field polarization. We approximate the external laser
field by a monochromatic wave, i.e.,
\begin{equation}
\mathbf{E}(t)=\varepsilon _{0}\sin \omega t\mathbf{e}_{x}.
\end{equation}%
This is a reasonable approximation for pulses whose duration is of
the order of ten cycles or longer (see, e.g. \cite{Carla3} for a
more detailed discussion). In this case, the electron momentum
distributions, when integrated over the transverse momentum
components, read
\begin{eqnarray}
F(p_{1\parallel },p_{2\parallel }) &=&\hspace*{-0.1cm}\iint \hspace*{-0.1cm}%
d^{2}p_{1\perp }d^{2}p_{2\perp }|M_{R}(\mathbf{p}_{1},\mathbf{p}_{2}) \\
&+&M_{L}(\mathbf{p}_{1},\mathbf{p}_{2})+\mathbf{p}_{1}\leftrightarrow
\mathbf{p}_{2}|^{2},  \notag  \label{distr}
\end{eqnarray}%
where $M_{R}(\mathbf{p}_{1},\mathbf{p}_{2})$ is given by Eq.~(\ref{Mp}) and $%
d^{2}p_{n\perp }=$ $p_{n\perp }dp_{n\perp }d\varphi _{p_{n}}$. If $\mathbf{A}%
(t\pm T/2)=-\mathbf{A}(t)$, where $T=2\pi /\omega $ denotes a field cycle,
the actions $S_{L}$, $S_{R}$ corresponding to the transition amplitudes $%
M_{L}$ and $M_{R\text{ }}$ obey the symmetry $S_{L}(\mathbf{p}_{1},\mathbf{p}%
_{2},t,t^{\prime },t^{\prime \prime })=S_{R}(\mathbf{-p}_{1},-\mathbf{p}%
_{2},t\pm T/2,t^{\prime }\pm T/2,t^{\prime \prime }\pm T/2).$ The
distributions have also been symmetrized with respect to the exchange $%
\mathbf{p}_{1}\leftrightarrow \mathbf{p}_{2}.$ To a good approximation, the
quantum-interference terms $\left[ M_{\nu }(\mathbf{p}_{1},\mathbf{p}_{2})%
\right] ^{\ast }M_{\mu }(\mathbf{p}_{1},\mathbf{p}_{2}),\nu \neq \mu ,$ get
washed out upon the transverse-momentum integration, so that it is
sufficient to add the above-stated amplitudes incoherently. Here, we
considered $V_{\mathbf{k}g}$ as constant and we integrate the transition
amplitude over the azimuthal angles $\varphi _{p_{n}}$.

We will now briefly discuss how the prefactors $V_{\mathbf{p}_{2}e}$ and $V_{%
\mathbf{p}_{1}e,\mathbf{k}g}$ behave with regard to the integration over $%
\varphi _{p_{n}}$. Obviously, if the second electron is excited from an $s$
state to an $s$ state, the prefactors $V_{\mathbf{p}_{2}e}$ and $V_{\mathbf{p%
}_{1}e,\mathbf{k}g}$ do not depend on this parameter. However, if a
transition from or to a $p$ state is considered there will be an angular
dependence in such prefactors.

For instance, tunneling ionization from a $p$ state would lead to the
argument $\beta (\mathbf{\tilde{p}}_{2}(t))$ in Eq. (\ref{Vp2e1s2p}). If
excitation from an $s$ state to a $p$ state or vice versa takes place, the
angular dependence of the prefactor $V_{\mathbf{p}_{1}e,\mathbf{k}g}$ is
given by $\beta (\mathbf{\kappa })$ in Eq. (\ref{Vp1k1s2p})). When
integrated over the azimuthal angles $\varphi _{p_{n}}$, $\left\vert \beta (%
\mathbf{\kappa })\right\vert ^{2}$ and $\left\vert \beta (\mathbf{p}%
_{2})\right\vert ^{2}$ will yield $2\pi ,$ so that the angular dependence of
these prefactors can be neglected. If, however, states of higher orbital
quantum numbers are involved, or if the initial and excited bound states of
the second electron are $p$ states, this dependence will be more complex.

The simplest scenario is if all prefactors are nonsingular, such as
in the velocity-gauge formulation of the SFA. In this case, they
will contribute to
the electron momentum distributions as $\left\vert V_{\mathbf{p}%
_{2}e}\right\vert ^{2}$ and $|V_{\mathbf{p}_{1}e,\mathbf{k}g}|^{2}.$ In the
length-gauge SFA, however, $V_{\mathbf{p}_{2}e}$ exhibits a singularity, and
therefore must be incorporated in the action according to Eq. (\ref%
{saddlemod}). This singularity, however, is present only in the radial part
of this prefactor. Therefore, the angular parts are still slowly varying,
and may be treated as above. The radial part of the prefactor, however, must
be incorporated in the action.

\subsection{Intensity dependence}

We will commence by having a closer look at how the momentum-space
constraints affect the electron momentum distributions for different
driving-field intensities. For that purpose, we will assume that the
prefactors $V_{\mathbf{p}_{1}e,\mathbf{k}g}$ and
$V_{\mathbf{p}_{2}e}$ are constant, and vary the laser-field
intensity. For the lowest intensity, the kinetic energy of the
returning electron is just enough to promote the second electron to
an excited state, i.e., we are considering the recollision
excitation (RESI) threshold $E_{2g}-E_{2e}\simeq 3.17U_{p}.$ This
intensity, however, is below the electron-impact ionization
threshold, i.e., $E_{2g}>3.17U_{p}.$ The intermediate intensity has
been chosen such that $E_{2g}-E_{2e}<3.17U_{p}$, i.e., above the
threshold for recollision-excitation. Nevertheless, this intensity
is not sufficient to make the second electron overcome the
ionization potential and be freed by electron-impact ionization.
Finally, the highest driving-field intensity considered in this
section is far above the recollision-excitation threshold, and
slightly above the electron-impact ionization threshold. This
implies that rescattering is classically allowed for both physical
mechanisms. The computations in this section have been
performed for Helium, and the pertaining results are presented in Fig.~\ref%
{intensity}.
\begin{figure}[tbp]
\begin{center}
\noindent \includegraphics[width=9cm]{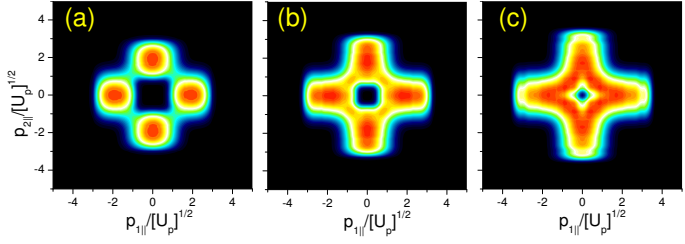}
\end{center}
\caption{Electron momentum distributions for Helium ($E_{1g}=0.97$ a.u., $%
E_{2g}=2$ a.u. and $E_{2e}=0.5$ a.u.) in a linearly polarized,
monochromatic field of frequency $\protect\omega =0.057$ a.u.. In
the picture, we considered all prefactors to be constant. Panels
(a), (b) and (c) correspond to a driving-field intensity
$I=2.16\times \mathrm{W/cm^{2}}$, $I=2.5\times \mathrm{W/cm^{2}}$
and $I=3\times \mathrm{W/cm^{2}}$, respectively. The contour plots
have been normalized to the maximum probability in each panel.}
\label{intensity}
\end{figure}

As an overall feature, the distributions exhibit four peaks at $%
(p_{j||},p_{\nu ||})=(\pm 2\sqrt{U_{p}},0),$ with $j,\nu =1,2$ and $j\neq
\nu .$ These peaks agree well with the constraints discussed in the previous
sections. This holds even if the driving-field intensity is just enough to
excite the second electron and the only allowed momenta are $\pm 2\sqrt{U_{p}%
}$ [Fig.~\ref{intensity}.(a)]. Physically, this means that the first
electron will reach its parent ion most probably at a crossing of the
driving field, reaching the detector with the most probable momenta of $\pm 2%
\sqrt{U_{p}},$ while the second electron will reach it with vanishing
momentum.

The shapes of the distributions, however, differ considerably. Indeed, at
the RESI threshold intensity [Fig.~\ref{intensity}.(a)], one observes
ring-shaped distributions. As the intensity increases, this distributions
become more and more elongated along the $p_{n\parallel }$ axis [Fig.~\ref%
{intensity}.(b)], until the maxima merge and cross-shaped
distributions are observed [Fig.~\ref{intensity}.(c)].

This change of shape may be understood by analyzing the
momentum-space constraints. The widths of the distributions are
determined by the tunnel ionization of the second electron from an
excited state. This process has no classical counterpart and leads
to distributions peaked at $p_{2\parallel }=0$ and which vanish at
$p_{2\parallel }=\pm 2\sqrt{U_{p}}$, i.e., at the direct ATI cutoff.
Increasing the intensity will only make the effective potential
barrier smaller or wider, and thus affect the overall yield, but
will not change such constraints.

The elongations in the distributions are determined by the rescattering of
the first electron. This rescattering, in contrast, delimits a momentum
region which is highly dependent on the driving-field intensity. Therefore,
its width in momentum space will vary. Specifically, at the RESI threshold,
there will be maxima in the distributions at $p_{1\parallel }=\pm 2\sqrt{%
U_{p}}$ due to the fact that the first electron rescatters most
probably at a field crossing. However, as these are the only
classically allowed momenta, the distributions will be fairly narrow
around this value. With increasing driving-field intensity, the
classically allowed region defined by Eq. (\ref{constraintp1}) will
become more and more extensive and this will cause the elongation.

Note that the electrons are indistinguishable so that the
above-stated arguments hold upon the exchange
$p_{1\parallel}\leftrightarrow p_{2\parallel}$. Hence, the
horizontal and vertical axis in the parallel momentum plane will be
equally affected.
\subsection{Bound-state signatures}

We will now investigate how the shape of the bound state to which the second
electron is excited is imprinted on the electron momentum distributions. We
will also employ different gauges and types of electron-electron
interaction. Explicitly, we will assume that the second electron is either
excited by a contact-type interaction $V_{12}^{(\delta )}(\mathbf{r}_{1}-%
\mathbf{r}_{2})=\delta (\mathbf{r}_{1}-\mathbf{r}_{2})$ or by a long-range,
Coulomb type interaction $V_{12}^{(C)}(\mathbf{r}_{1}-\mathbf{r}_{2})=1/(%
\mathbf{r}_{1}-\mathbf{r}_{2})$. In the former case, $V_{12}^{(\delta )}(%
\mathbf{p}_{1}-\mathbf{k})=const.$, while in the latter case $V_{12}^{(C)}(%
\mathbf{p}_{1}-\mathbf{k})\sim 1/(\mathbf{p}_{1}-\mathbf{k})^{2}.$

In order to perform a direct comparison, we will take the same parameters as
in Fig.~\ref{intensity}, but incorporate the prefactors $V_{\mathbf{p}_{1}e%
\mathbf{,k}g}^{(1s\rightarrow 2s)}$ and $V_{\mathbf{p}_{2}e}^{(2s)}$, or $V_{%
\mathbf{p}_{1}e\mathbf{,k}g}^{(1s\rightarrow 2p)}$ and $V_{\mathbf{p}%
_{2}e}^{(2p)}$, corresponding to the $1s\rightarrow 2s$ or $1s\rightarrow 2p$
excitation with subsequent tunneling, respectively.

\begin{figure}[tbp]
\begin{center}
\noindent \includegraphics[width=9cm]{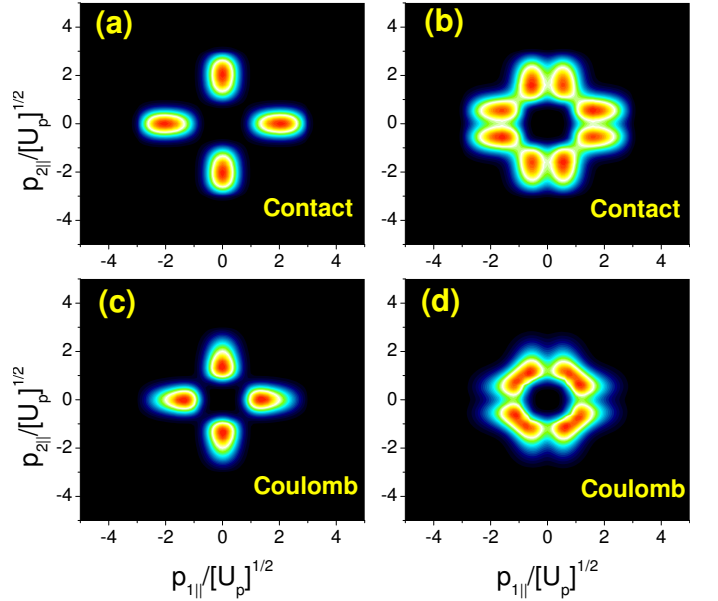}
\end{center}
\caption{Velocity-gauge electron momentum distributions for Helium ($%
E_{1g}=0.97$ a.u., $E_{2g}=2$ a.u. and $E_{2e}=0.5$ a.u.) in a
linearly polarized, monochromatic field of frequency $\protect\omega
=0.057$ a.u. and intensity $I=2.16\times \mathrm{W/cm^{2}}$. In
panels (a) and (c), the first electron has been excited to $2s$,
while in panels (b), and (d) it has been excited to $2p$. The
interaction employed is indicated in the figure. The contour plots
have been normalized to the maximum probability in each panel.}
\label{rings}
\end{figure}

In Fig.~\ref{rings}, we consider the lowest intensity in the previous figure
and the velocity gauge. If the electron is excited to the $2s$ state [Fig.~%
\ref{rings}.(a)], we observe four spots which are slightly elongated along
the $p_{n\parallel }$ axis. Hence, in comparison to its constant prefactor
counterpart, i.e.,Fig.~\ref{intensity}.(a), there was a narrowing. This
narrowing is caused by the interplay of two features in the prefactor $V_{%
\mathbf{p}_{2}e}^{(2s)}$. First, this prefactor exhibits two symmetric
nodes, which, for vanishing transverse momentum are located at $%
p_{2\parallel }=\pm \sqrt{U_{p}}$. As the transverse momentum increases,
these minima move towards vanishing parallel momenta. Second, $V_{\mathbf{p}%
_{2}e}^{(2s)}$ decreases very steeply with transverse momenta $\mathbf{p}%
_{2\perp }$. Hence, upon integration over this parameter, the main
contributions will be caused by small values of $\mathbf{p}_{2\perp }$ and
will vanish near $p_{2\parallel }=\pm \sqrt{U_{p}}$.

If, on the other hand, one assumes that the second electron is excited to $%
2p,$ there is both a broadening in the distributions and a splitting in
their peaks. These features are depicted in Fig.~\ref{rings}.(b). The
splitting occurs at the axis $p_{n\parallel }=0,n=1,2,$ and is caused by the
fact that $V_{\mathbf{p}_{2}e}$ exhibits a very pronounced node at vanishing
momenta, i.e., exactly where one expects $\mathrm{Im}[t]$ to be minimum and
the yield to be maximum. This has been verified by a direct inspection of
the radial dependence of Eq.~(\ref{Vp2e1s2p}), and omitting the $V_{\mathbf{p}%
_{2}e}$ prefactor in our computations. The latter procedure caused the
additional minima to disappear (not shown). The broadening in the
distributions as compared to the $1s\rightarrow 2s$ case is a consequence of
the much slower decrease in $V_{\mathbf{p}_{2}e}^{(2p)}$ with increasing
transverse momentum $p_{2\perp }$ and of the absence of the nodes at $%
p_{2\parallel }=\pm \sqrt{U_{p}}$. There are also additional nodes at the
diagonal $p_{1||}=p_{2||}$ and at the anti-diagonal $p_{1||}=-p_{2||}$ of
the $p_{1||}p_{2||}$ plane.

In this intensity regime, there seems to be little difference in the shapes
of the distributions if the electron is excited to the $2s$ state,
regardless of whether the first electron interacts with its parent ion
through a contact or a Coulomb interaction [Figs. \ref{rings}.(a) and (c),
respectively]. This is possibly caused by the fact that the prefactor $V_{%
\mathbf{p}_{2}e}^{(2s)}$, due to its fast-decaying behavior, delimits a very
narrow region in momentum space. This adds up to the very restrictive
momentum constraints. In contrast, the effect of the Coulomb tail is much
more critical if the electron is promoted to the $2p$ state. Indeed, for a
Coulomb type interaction [Fig. \ref{rings}.(d)], the splitting of the peaks
at the axis $p_{n\parallel }=0$ remain, but the nodes at $p_{1||}=p_{2||}$
and $p_{1||}=-p_{2||}$ disappear as compared to its contact-interaction
counterpart [Fig. \ref{rings}.(b)]. This is caused by the fact that the
former minima are a characteristic of the $V_{\mathbf{p}_{2}e}^{(2p)}$
prefactor, whereas the latter are mainly determined by momentum-space
effects. The Coulomb interaction introduces a further momentum bias, and
washes out the latter nodes.

\begin{figure}[tbp]
\begin{center}
\noindent \includegraphics[width=9cm]{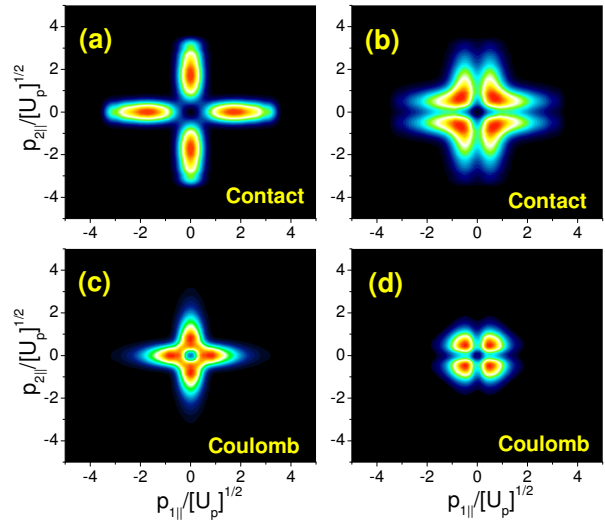}
\end{center}
\caption{Velocity-gauge electron momentum distributions for the same
parameters as in the previous figure, but driving-field intensity
$I=3\times \mathrm{W/cm^{2}}$. In panels (a) and (c), the first
electron has been excited to $2s$, while in panels (b), and (d) it
has been excited to $2p$. The interaction employed is indicated in
the figure. The contour plots have been normalized to the maximum
probability in each panel.} \label{Heliumhigher}
\end{figure}

We will now discuss what happens if the intensity of the driving field is
such that $3.17U_{p}>E_{2g}-E_{2e}.$ These results are displayed in Fig.~\ref%
{Heliumhigher}, for the highest intensity in Fig.~\ref{intensity}. As
expected, all distributions are much more elongated along the axis $%
p_{n\parallel }=0$, as compared to the low-intensity case. The imprint,
however, of the different bound states to which the second electron is
excited and from which it subsequently tunnels are the same as in the
below-threshold regime. Indeed, we notice that there is a narrowing in the
distributions for the $1s\rightarrow 2s$ case [Figs.~\ref{Heliumhigher}.(a)
and (c)], and a splitting in the peaks at the four axis in the $%
1s\rightarrow 2p$ case [Figs.~\ref{Heliumhigher}.(b) and (d)]. This is not
surprising, as the prefactors $V_{\mathbf{p}_{2}e}^{(2s)}$ and $V_{\mathbf{p}%
_{2}e}^{(2p)}$ exhibit the same functional dependencies as before.

The shapes of distributions, however, change much more critically in
this intensity regime, with regard to the type of electron-electron
interaction, than for the intensity used in Fig.~\ref{rings}. For
all cases, the distributions computed using the Coulomb-type
interaction (Figs.~\ref{Heliumhigher}.(c) and
~\ref{Heliumhigher}.(d)) are much more localized in the low-momentum
regions than those computed with a contact-type interaction (see
Figs.~\ref{Heliumhigher}.(a) and ~\ref{Heliumhigher}.(b)). This is
expected, as  $V_{12}(\mathbf{p}_{1}-\mathbf{k})$ favors low momenta
for the former, while it is constant for the latter.  Physically,
this reflects the fact that rescattering of the first electron is
now allowed to occur over an extensive region in momentum space.
Hence, it does make a difference whether the second electron is
excited by a long-range or zero-range interaction.

We will now perform an analysis of the electron-momentum
distributions in the length gauge. In this case, the prefactor
$V_{\mathbf{p}_{2}e}$ governing the tunneling of the second electron
exhibits a singularity, and must be incorporated in the action. The
modifications in the action read, for the $2s$ and $2p$ bound
states,
\begin{equation}
-i\partial _{t}\ln V_{\mathbf{p}_{2}e}^{(2s)}=-i\frac{2E(t)\tilde{p}%
_{2}(t)_{\parallel }(\left[ \tilde{p}_{2}(t)_{\parallel }\right] ^{2}+%
\mathbf{p}_{2\perp }^{2}-6E_{2e})}{\chi _{+}(\mathbf{\tilde{p}}_{2}(t))\chi
_{-}(\mathbf{\tilde{p}}_{2}(t))},
\end{equation}%
and
\begin{equation}
-i\partial _{t}\ln \tilde{V}_{\mathbf{p}_{2}e}^{(2p)}=i\frac{E(t)(\mathbf{p}%
_{2\perp }^{2}+2E_{2e}-3\left[ \tilde{p}_{2}(t)_{\parallel }\right] ^{2})}{%
\tilde{p}_{2}(t)_{\parallel }\left( \left[ \tilde{p}_{2}(t)_{\parallel }%
\right] ^{2}+\mathbf{p}_{2\perp }^{2}+2E_{2e}\right) },
\end{equation}
respectively, with \newline $\chi _{\pm }(\mathbf{\tilde{p}}_{2}(t))=\left( \left[ \tilde{p}%
_{2}(t)_{\parallel }\right] ^{2}+\mathbf{p}_{2\perp }^{2}\pm
2E_{2e}\right)$ and $E(t)=-\partial _{t}A(t)$.

For each orbit, the tunneling time of the second electron will split
into two values, as compared to the non-modified action. This has
particularly important consequences as far as $\mathrm{Im}[t]$ is
concerned, since it provides a rough measure of the width of the
barrier through which the second electron tunnels. Physically, this
means there will be one set of orbits for which the effective
potential barrier will be widened, and another one for which it will
be narrowed.

\begin{figure}[tbp]
\begin{center}
\noindent \includegraphics[width=9cm]{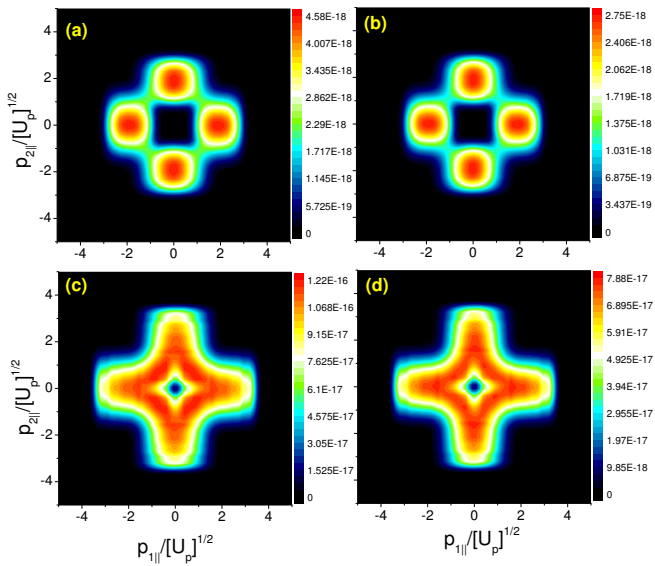}
\end{center}
\caption{Length-gauge electron momentum distributions for Helium in a
linearly polarized, monochromatic field. Throughout, we assumed $V_{\mathbf{p%
}_1e,\mathbf{k}g}=const.$, $V_{12}$ to be a contact-type interaction, and
incorporated $V^{(2s)}_{\mathbf{p}_2e}$ in the action. Panels (a) and (c)
correspond to the trajectories for which the barrier has been narrowed by
the modifications in the action, while Panels (b) and (d) correspond to
those for which it has been widened. The upper and lower panels correspond
to driving field intensities $I=2.16 \times 10^{14}\mathrm{W}/\mathrm{cm}^2$
and $I=3 \times 10^{14}\mathrm{W}/\mathrm{cm}^2$, respectively. In order to
perform a quantitative comparison, we are providing the explicit values for
the NSDI yield.}
\label{length1}
\end{figure}

In Fig.~\ref{length1}, we present the contributions of each of the orbits
resulting from this splitting for a final $2s$ state, for two different
driving-field intensities. For simplicity, in order to single out the effect
of the modified action, we took the rescattering prefactor $V_{\mathbf{p}_1e,%
\mathbf{k}g}$ to be constant.

In general, the distributions differ quantitatively in a factor between $1.5$
and $1.7$, depending on whether $\mathrm{Im}[t]$ decreased [Figs.~\ref%
{length1}(a) and (c)] or increased [Figs.~\ref{length1}(b) and (d)]. This
shows that the splitting in this quantity is small, and therefore both
contributions are comparable.

Furthermore, the distributions displayed in Fig.~\ref{length1} are
strikingly similar to those observed in Fig.~\ref{intensity} (see panels (a)
and (c) therein), for which only constant prefactors have been considered.
Indeed, the width of all distributions, along the axis, is determined by the
direct ATI cutoff, i.e., $-2\sqrt{U_{p}}\leq p_{n\parallel }\leq 2\sqrt{U_{p}%
}$. At first sight, this is unexpected, as we are assuming that the second
electron is tunneling from a $2s$ state. As previously discussed, the
prefactor $V_{\mathbf{p}_{2}e}^{(2s)}$ exhibits a node in $p_{2\parallel
}=\pm \sqrt{U_{p}}$, which leads to a narrowing of the distributions along
the $p_{n\parallel }$ axis. An inspection of Eq. (\ref{Vp2e2s}) also
suggests that, were it not for its singularity, the length-gauge prefactor
would be very similar to the velocity-gauge prefactor. This is a consequence
of the fact that the second electron is leaving when the field $E(t)$ is
near its maximum. For a monochromatic field, this implies that the vector
potential $A(t)$ is practically vanishing.

One should note, however, that we are considering only the
individual contributions from each of the orbits originating from
the modification of the action. It is very likely that, in order to
recover the structure determined by the prefactor
$V^{(2s)}_{\mathbf{p}_{2}e}$, one must consider the coherent
superposition of all the orbits originating from the splitting of
$\mathrm{Im}[t]$ when computing the yield. Since these contributions
are comparable, one expects the above-mentioned nodes to be
recovered due to quantum-interference effects.

\subsection{Comparison with experiments}

We will now perform a direct comparison with the results in Ref.~\cite%
{Eremina}. In particular, in this reference, the distributions encountered
have been modeled employing the electron-impact ionization physical
mechanism and a modified ionization threshold for the second electron. Apart
from that, however, in view of the driving-field intensities involved, one
expects recollision-excitation tunneling to be present.

For that purpose, we will consider argon and the same laser-field
parameters as in Ref.~\cite{Eremina} (c.f. Fig.~2 therein). We will
assume, however, that, when the first electron recollides, it
excites the second electron from the $3p$ state either to the $4s$
or to the $4p$ state. Thereby, we took the velocity gauge, and
assumed that the first electron interacts with the ion by a Coulomb
or contact interaction.
\begin{figure}[tbp]
\begin{center}
\noindent \includegraphics[width=8cm]{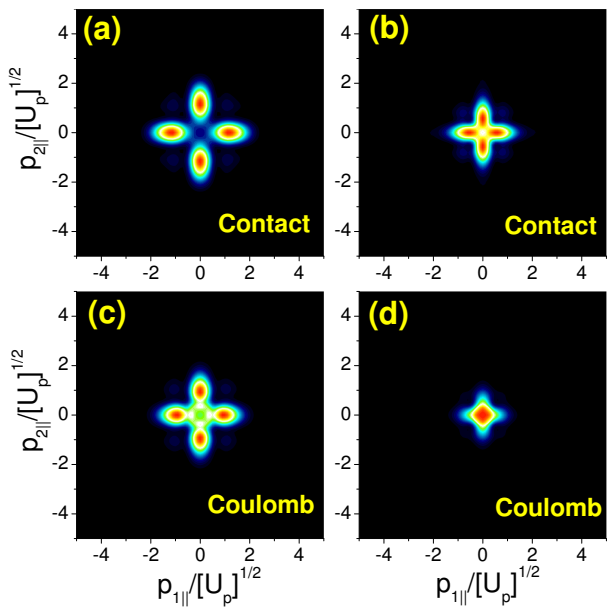}
\end{center}
\caption{Velocity-gauge electron momentum distributions for Argon in
a linearly polarized, monochromatic field of frequency
$\protect\omega=0.057$ a.u. The electron is excited from $3p$ to
$4s$, i.e., $E_{1g}=0.58$ a.u., $E_{2g}=1.02$ a.u. and $E_{2e}=0.40
$ a.u. in our calculations. The laser-field intensity in panels (a)
and (c), and panels (b) and (d) is $I=9 \times
10^{13}\mathrm{W}/\mathrm{cm}^2$ and $I=1.5 \times
10^{14}\mathrm{W}/\mathrm{cm}^2$, respectively. The type of
interaction $V_{12}$ taken is indicated in the figure. The contour
plots have been normalized to the maximum probability in each panel.
We have verified, however, that the highest yields on left-hand
panels are between one and a half and two orders of magnitude
smaller than those on the right-hand side.} \label{argon}
\end{figure}

The results for the $3p\rightarrow4s$ excitation are presented in Fig.~\ref%
{argon}. An overall feature in the distributions are two main maxima along
the $p_{n\parallel}$, $n=1,2$, axis. These features are mainly caused by the $%
V^{(4s)}_{\mathbf{p}_2e}$ prefactor for the tunnel ionization of the
second electron, which decays very rapidly with increasing
transverse momenta and exhibit nodes near $p_{2\parallel}=\pm
0.5\sqrt{U_p}$. In general, we have verified that this prefactor
determines the shape of the electron-momentum distributions.
Secondary maxima, around one order of magnitude smaller,
occur due to the rescattering prefactor $V^{(3p \rightarrow 4s)}_{\mathbf{p}%
_1e,\mathbf{k}g}$. This prefactor exhibits an annular shape around $%
p_{1\parallel}=p_{2\parallel}=0$.

The existing experiments, however, do not lead to distributions concentrated
along the axis of the $p_{1\parallel}p_{2\parallel}$ plane. The results for
Helium in the previous section suggest that a $p$ state may lead to broader
distributions. For that reason, we will assume that, instead, the second
electron is excited to the $4p$ state.

\begin{figure}[tbp]
\begin{center}
\noindent \includegraphics[width=8cm]{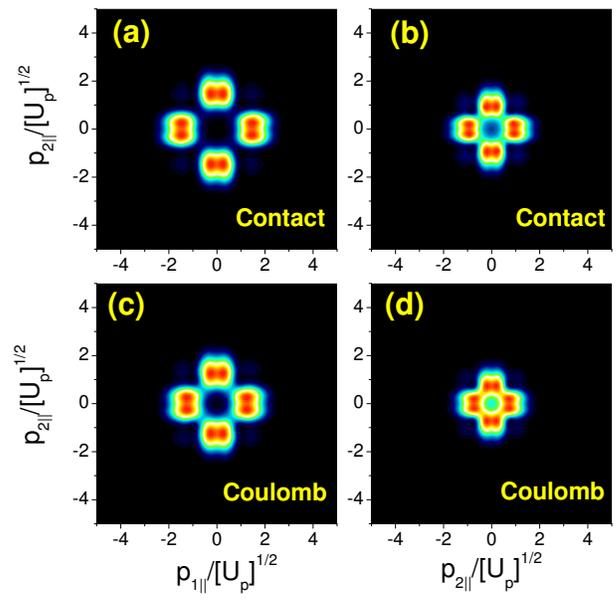}
\end{center}
\caption{Velocity-gauge electron momentum distributions for Argon in
a linearly polarized, monochromatic field of frequency
$\protect\omega=0.057$ a.u. The electron is excited from $3p$ to
$4p$, i.e., $E_{1g}=0.58$ a.u., $E_{2g}=1.02$ a.u. and $E_{2e}=0.31
$ a.u. in our calculations. The laser-field intensity in panels (a)
and (c), and panels (b) and (d) is $I=9 \times
10^{13}\mathrm{W}/\mathrm{cm}^2$ and $I=1.5 \times
10^{14}\mathrm{W}/\mathrm{cm}^2$, respectively. The type of
interaction $V_{12}$ taken is indicated in the figure. The contour
plots have been normalized to the maximum probability in each panel.
We have verified, however, that the highest yields on left-hand
panels are between one and two orders of magnitude smaller than
those on the right-hand side.} \label{4pArall}
\end{figure}

Fig.~\ref{4pArall} depicts the electron-momentum distributions
for Argon under the assumption that the electron was excited from $3p$ to $%
4p $. All distributions in the figure exhibit four main maxima,
which are broader than those in Fig.~\ref{argon} and almost split at
the axis $p_{n\parallel}=0$. These maxima are mainly determined by
the prefactor  $V^{(4p)}_{\mathbf{p}_2e}$, which has a node at the
axis for low transverse momenta and nodes around $p_{2\parallel}=\pm
\sqrt{U_p}$ across a wide transverse-momentum range.
Apart from that, the prefactors $V^{(3p \rightarrow 4p)}_{%
\mathbf{p}_1e,\mathbf{k}g}$ decay more slowly with regard to the
transverse momenta. This implies that, upon integration, a larger
momentum region will be contributing to the NSDI yields. As in the
previous case, this prefactor also leads to secondary maxima (see
Figs.~\ref{4pArall}.(c) and (d) for concrete examples). In all
cases, both in Figs.~\ref{argon} and \ref{4pArall}, a Coulomb type
interaction mainly introduces a bias towards lower momenta.

\begin{figure}[tbp]
\begin{center}
\noindent \includegraphics[width=8cm]{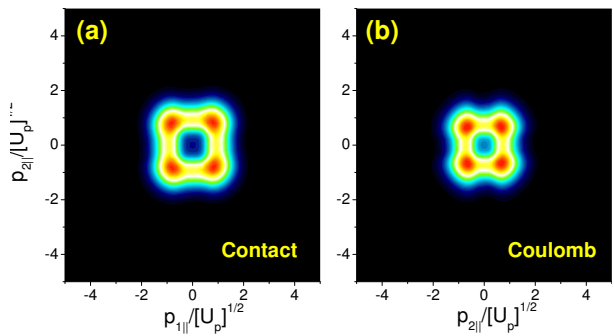}
\end{center}
\caption{Velocity-gauge electron momentum distributions for Argon in
a linearly polarized, monochromatic field of frequency
$\protect\omega=0.057$ a.u. and intensity $I=1.5 \times
10^{14}\mathrm{W}/\mathrm{cm}^2$. The electron is excited from $3p$
to $4p$. We have taken the prefactor $V_{\mathbf{p}_2,e}$ to be
constant. The type of interaction $V_{12}$ is indicated in the
figure. The contour plots have been normalized to the maximum
probability in each panel.} \label{Ar3p4p}
\end{figure}

Despite the above-mentioned broadening, the electron-momentum
distributions in Fig.~\ref{4pArall} are still considerably narrower
than those observed in Ref. \cite{Eremina}. Within our framework,
this constraint is imposed by the $V^{(4p)}_{\mathbf{p}_2e}$
prefactor. In fact, we have verified that, for large principal
quantum number, this prefactor always exhibits nodes at lower
absolute momenta than the ATI cutoff of
$p_{2\parallel}=2\sqrt{U_p}$. In fact, if $V_{\mathbf{p}_2e}$ is
taken to be constant, the distributions become considerably broader
and a better agreement with the experiments is obtained. This is
shown in Fig.~\ref{Ar3p4p}, as ring-shaped distributions with four
symmetric maxima at $p_{1\parallel}=p_{2\parallel}$ and
$p_{1\parallel}=-p_{2\parallel}$. Such maxima are mainly determined
by the $V^{(3p \rightarrow 4p)}_{%
\mathbf{p}_1e,\mathbf{k}g}$ prefactor. One should note, however,
that this procedure is inconsistent from a theoretical perspective:
Since the electron has been excited to the $4p$ state, it should
subsequently tunnel from it. Hence, the pertaining prefactor must be
taken.

\section{Conclusions}
\label{conclusions}

Our analysis of the rescattering-excitation ionization (RESI)
mechanism shows that the NSDI electron momentum distributions depend
on the interplay between the relevant momentum-space regions, the
type of interaction exciting the second electron, and the spatial
dependence of the bound states involved. We will commence by
discussing each of these issues separately.

The shapes of the electron momentum distributions, are determined by
the interplay between two different behaviors, associated with the
collision of the first electron and the tunneling of the second
electron. The momentum region determined by the tunnel ionization of
the second electron from an excited state will be always restricted
by the direct ATI cutoff. The relevant momentum region will not
change regardless of the driving-field intensity, as this will
always be a classically forbidden process.

The first electron, on the other hand, rescatters inelastically with
its parent ion, giving part of its kinetic energy upon return to
excite the second electron. Hence, if its maximum return energy is
larger than the energy difference $E_{2g}-E_{2e}$, rescattering has
a classical counterpart. This implies that there will be a
classically allowed region in momentum space. If, however, this
energy is just enough to excite the second electron, the classical
region will collapse. Hence, the extension of the relevant region in
momentum space related to the rescattering of the first electron
will depend on the driving-field intensity. Hence, the distributions
become increasingly elongated as the intensity increases.

This also implies that one may define a threshold driving-field
intensity for the RESI mechanism. This intensity is considerably
lower than that necessary for the second ionization potential to be
overcome by the second electron, i.e., for electron-impact
ionization to occur.

Apart from that, we have observed that the bound states involved in
the process leave very distinct fingerprints on the electron
momentum distributions. This is particularly true for the bound
state of the second electron, prior and subsequently to excitation.
In fact, the widths of the distributions, their shapes and the
number of maxima present will strongly depend on the principal and
orbital quantum numbers of the bound states involved.

In contrast, the type of interaction $V_{12}$ by which the second
electron is excited influences such distributions in a less drastic
way. Indeed, a long-range, Coulomb interaction mainly introduces a
bias towards lower momenta, as compared to a contact-type
interaction.

A very important observation is that all distributions encountered
in this work are equally spread over the four quadrants of the
$p_{1\parallel}p_{2\parallel}$ plane. Under no circumstances have we
found electron momentum distributions concentrated only on the
second and fourth quadrant of this plane, as reported in the
literature \cite{timelag,NSDIalign,Chen}.

Within our framework, the above-stated symmetry can immediately be
inferred from Eq. (\ref{distr}). Nonetheless, one could argue that
our approach does not include the residual binding potential in the
electron propagation in the continuum. Recent results, however, from
a classical-trajectory computation in which the Coulomb potential
has been incorporated, also revealed the same symmetry if only the
RESI mechanism is singled out \cite{Agapi09}. This is a strong hint
that our results are not an artifact of the strong-field
approximation.

Hence, we suspect that, in the existing literature, the
contributions from the RESI mechanism to nonsequential double
ionization  also equally occupy the four quadrants of the
$p_{1\parallel}p_{2\parallel}$ plane. They may, however, be
difficult to extract, as explained below.

In many situations addressed in the literature, the driving-field
intensity is high enough for electron-impact ionization to occur.
This means that this latter NSDI mechanism is also present, and
fills the first and third quadrant of the
$p_{1\parallel}p_{2\parallel}$ plane. Since, in many ab initio
 models, the different rescattering mechanisms are
difficult to disentangle, the contributions from electron-impact
ionization possibly obscure those from RESI in this region. In the
second and fourth quadrant of the parallel momentum plane, the
former contributions are absent and those from RESI can be more
easily identified. In our approach, electron-impact ionization is
absent from the start.

If the driving-field intensities are below the electron-impact
ionization threshold, the second electron may no longer be provided
with enough energy to overcome the second ionization potential.
Consequently, RESI becomes more prominent and the distributions
equally occupy the four quadrants of the parallel momentum plane. In
fact, ring-shaped distributions centered around
$p_{1\parallel}=p_{2\parallel}=0$ have been observed experimentally
for this intensity region \cite{threshold,Eremina}.

Our results are far more localized near the $p_{n\parallel}=0$ axis
than the experimental findings. This discrepancy may be due to the
following reasons. First, for higher intensities employed in Ref.
\cite{Eremina}, collisional excitation may take place not only to
the $4s$ or to the $4p$ state, but also to highly lying states, or
to a coherent superposition of excited states. To take this into
account may be needed in order to reproduce the experimental data.

Second, at the relevant driving-field intensities, one expects the
excited states to be distorted by the field, and the propagation of
the electron in the continuum near the core to be influenced by the
residual ionic potential. This implies that a semi-analytical
treatment beyond the strong-field approximation is necessary (see,
e.g., \cite{corrections,Denys} for other phenomena and the
electron-impact ionization case, respectively). Such a treatment is
outside the scope of the present paper, and will be the topic of
future work.

\vspace*{0.5cm}
 \noindent \textbf{Acknowledgements:} This work
has been financed by the UK EPSRC (Grant no. EP/D07309X/1). We would
also like to thank the UK EPSRC for the provision of a DTA
studentship and of a summer studentship, and A. Emmanouilidou for
providing her RESI results prior to publication. C.F.M.F. and T.S.
are also grateful to the ICFO Barcelona for its kind hospitality.

\section{Appendix}

In this appendix, we provide the general expressions for all
prefactors employed in this paper. We will make no simplifying
assumption on the initial states of the first and second electron,
and on the excited state to which the second electron is promoted,
apart from the fact that they are given by hydrogenic wavefunctions.
In order to compute the prefactors, we will employ the expansion
\begin{eqnarray}
e^{-i\mathbf{q}\cdot \mathbf{r}_{\alpha }} &=&4\pi \sum_{l=0}^{\infty
}\sum_{m=-l}^{l}(-i)^{l}j_{l}(qr_{\alpha })  \notag \\
&&Y_{l}^{m}(\theta _{q_{\alpha }},\varphi _{q_{\alpha }})\left[ Y_{l^{\prime
}}^{m^{\prime }}(\theta _{\alpha },\varphi _{\alpha })\right] ^{\ast },
\label{planeexpansion}
\end{eqnarray}%
where $\mathbf{q}$ denotes a generic momentum, $\mathbf{r}_{\alpha }$ the
coordinate of the ${\alpha }^{th}$ electron, and $j_{l}(\cdot )$ the
spherical Bessel functions of the first kind. This expression will be both
used in the derivation of $V_{\mathbf{p}_{2}e}$ and $V_{\mathbf{p}_{1}e%
\mathbf{,k}g},$ together with the orthogonality relation
\begin{equation}
\int \left[ Y_{l^{\prime }}^{m^{\prime }}(\theta _{\alpha },\varphi _{\alpha
})\right] ^{\ast }Y_{l}^{m}(\theta _{\alpha },\varphi _{\alpha })d\Omega
=\delta _{ll^{\prime }}\delta _{mm^{\prime }},
\end{equation}%
where $\Omega $ denotes the solid angle.

For the former prefactor, Eq. (\ref{Vp2e}) reduces to
\begin{eqnarray}
V_{\mathbf{p}_{2}e} &\sim & \int \hspace*{-0.2cm}\int \exp [-i(\mathbf{%
\tilde{p}}_{2}(t)\cdot \mathbf{r}_{2})]R_{nl}(r_{2})Y_{l}^{m}(\theta
_{2},\varphi _{2})r_{2}dr_{2}d\Omega  \notag \\
&=& 4\pi (-i)^{l}Y_{l}^{m}(\theta _{\tilde{p}_{2}},\varphi _{\tilde{p}_{2}})%
\mathcal{I}_{1},
\end{eqnarray}%
with
\begin{equation}
\mathcal{I}_{1}=\int_{0}^{\infty }\hspace*{-0.35cm}r_{2}R_{nl}(r_{2})j_{l}(%
\tilde{p}_{2}(t)r_{2})dr_{2}.
\end{equation}%
Similarly, Eq. (\ref{resc1st}) reads
\begin{equation}
V_{\mathbf{p}_{1}e\mathbf{,k}g}\sim V_{12}(\mathbf{p}_{1}-\mathbf{k})%
\mathcal{I}_{2},
\end{equation}%
with
\begin{eqnarray}
\mathcal{I}_{2} &=&\int d^{3}r_{2}e^{i(\mathbf{k-p}_{1})\cdot \mathbf{r}%
_{2}}R_{n_{e}l_{e}}(r_{2})\left[ Y_{l_{e}}^{m_{e}}(\theta _{2},\varphi _{2})%
\right] ^{\ast }  \notag \\
&&R_{n_{g}l_{g}}(r_{2})Y_{l_{g}}^{m_{g}}(\theta _{2},\varphi _{2}),
\label{angularintegral}
\end{eqnarray}%
where the indices $g$ and $e$ in the principal, orbital and magnetic quantum
numbers refer to the ground and excited states, respectively.

We will now compute the radial integrals $\mathcal{I}_{1}$ and $\mathcal{I}%
_{2}$ explicitly. For that purpose, let us consider a generic Hydrogenic
radial wavefunction
\begin{eqnarray}
R_{nl}(r) &=&C_{nl}r^{l}\exp [-\sqrt{2E_{n}}r]  \notag \\
&&\times \sum_{\nu =0}^{n-l-1}\frac{2^{\nu }(-1)^{\nu +1}(\sqrt{2E_{n}}%
)^{\nu }}{\nu !(n-l-1-\nu )!(2l+1+\nu )!},
\end{eqnarray}%
with
\begin{equation}
C_{nl}=-\left\{ \frac{(2\sqrt{2E_{n}})^{3+2l}(n-l-1)!}{2n\left[ (n+l)!\right]
^{3}}\right\} ^{1/2}\left[ (n+l)!\right] ^{2}.
\end{equation}%
In the above-stated equations, $E_{n}$ denotes the energy of the bound state
to be studied, i.e., $n=2g$ or $n=2e$ for the ground or excited states of
the second electron, respectively. Since we are performing a qualitative
analysis, we will concentrate mostly on the functional form of $R_{nl}(r).$
The integral $\mathcal{I}_{1}$ present in the prefactor $V_{\mathbf{p}_{2}e}$
can then be written as
\begin{eqnarray}
\mathcal{I}_{1}\hspace*{-0.2cm} &\varpropto &\sum_{\nu =0}^{n-l-1}\frac{%
(-1)^{\nu +1}2^{\nu -1-l}(\sqrt{2E_{n}})^{-2-l}}{\nu !(n-l-1-\nu )!(2l+1+\nu
)!}\frac{\Gamma \left[ 2+\nu +2l\right] }{\Gamma \left[ 3/2+l\right] }
\notag \\
&&\times _{2}F_{1}(1+l+\frac{\nu }{2},\frac{3+\nu }{2}+l,\frac{3}{2}+l,-%
\frac{\left[ \tilde{p}_{2}(t)\right] ^{2}}{2E_{e}}).
\end{eqnarray}

The integral $\mathcal{I}_{2}$ in $V_{\mathbf{p}_{1}e\mathbf{,k}g}$ is
slightly more involved. It may be explicitly written as
\begin{equation}
\mathcal{I}_{2}=4\pi \sum_{l=0}^{\infty
}\sum_{m=-l}^{l}(-i)^{l}Y_{l}^{m}(\theta _{q_{\alpha }},\varphi _{q_{\alpha
}})\mathcal{I}_{2R}\mathcal{I}_{2\Omega },
\end{equation}%
where
\begin{equation}
\mathcal{I}_{2R}=\int_{0}^{\infty
}r_{2}^{2}R_{n_{g}l_{g}}(r_{2})R_{n_{e}l_{e}}(r_{2})j_{l}(\kappa r_{2})dr_{2}
\end{equation}%
and
\begin{equation}
\mathcal{I}_{2\Omega }=\int Y_{l}^{m}(\theta _{2},\varphi _{2})\left[
Y_{l_{e}}^{m_{e}}(\theta _{2},\varphi _{2})\right] ^{\ast
}Y_{l_{g}}^{m_{g}}(\theta _{2},\varphi _{2})d\Omega
\end{equation}%
give the radial and angular dependencies of such prefactors, respectively.
The explicit expression for $\mathcal{I}_{2}$ is then%
\begin{eqnarray}
\mathcal{I}_{2} &\hspace*{-0.2cm}=\hspace*{-0.2cm}&4\pi \hspace*{-0.2cm}%
\sum_{l=|l_{g}-l_{e}|}^{l_{g}+l_{e}}\hspace*{-0.1cm}%
\sum_{m=-l}^{l}(-i)^{l}(-1)^{m_{e}}Y_{l}^{m}(\theta _{\kappa },\varphi
_{\kappa })\sqrt{\frac{\left( 2l_{g}+1\right) \left( 2l_{e}+1\right) }{4\pi
\left( 2l+1\right) }}  \notag \\
&&\times \left\langle l_{g},l_{e},0,0\right. \left\vert l,0\right\rangle
\left\langle l_{g},l_{e},m_{g},-m_{e}\right. \left\vert l,0\right\rangle
\mathcal{I}_{2R}.  \label{totalintegral}
\end{eqnarray}
The radial integral $\mathcal{I}_{2R}$ is proportional to
\begin{widetext}
\begin{eqnarray}
\mathcal{I}_{2R} &\propto &\sum_{\nu
_{g}=0}^{n_{g}-l_{g}-1}\sum_{\nu
_{e}=0}^{n_{e}-l_{e}-1}\frac{(-1)^{\nu _{e}+\nu _{g}+1}2^{\nu
_{e}+\nu _{g}-1-l}(\sqrt{2E_{2g}})^{\nu _{g}}(\sqrt{2E_{2e}})^{\nu
_{e}}\left[ \zeta (E_{2g},E_{2e})\right] ^{-3-\nu _{e}-\nu
_{g}-l_{e}-l_{g}}}{\nu _{e}!\nu _{g}!(n_{e}-l_{e}-1-\nu
_{e})!(n_{g}-l_{g}-1-\nu _{g})!(2l_{e}+1+\nu
_{e})!(2l_{g}+1+\nu _{g})!}  \notag \\
&&\frac{\Gamma \left[ 2+\lambda \right] }{\Gamma \left[ 3/2+l\right] }\left(
\frac{\kappa ^{2}}{\zeta ^{2}(E_{2g},E_{2e})}\right) ^{l/2}\hspace*{-0.1cm}%
_{2}F_{1}(\frac{3+\lambda }{2},\frac{4+\lambda }{2},\frac{3}{2}+l,-\frac{%
\kappa ^{2}}{\zeta ^{2}(E_{2g},E_{2e})}),
\end{eqnarray}%
\end{widetext}
where $\lambda =\nu _{e}+\nu _{g}+l_{e}+l_{g}+l$ and $\zeta (E_{2g},E_{2e})$
is defined according to Eq. (\ref{zeta}). Note that the terms in Eq. (\ref%
{totalintegral}) are only non-vanishing if $m=m_g-m_e$ and $l_1+l_2-l$ is
even.

In the present work, apart from the case in which only $s$ states are
involved and the angular integrals are constant, one may identify the
following cases. First, the second electron may be initially in a $p$ state
and be excited to an $s$ state. In this case, $l=l_g=1$ and $l_e=0$. Second,
if the electron is initially in an $s$ state and is excited to a $p$ state,
then $l=l_e=1$ and $l_g=0$. Finally, if the second electron suffers a
transition from a $p$ state to another $p$ state, in principle $l=0,1,2$.
Due to the constraints upon $l$ for the Clebsch-Gordan coefficients,
however, only the terms with $l=0,2$ will survive. Apart from that, the
constraint upon $m$ will impose further restrictions for $m_e$ and $m_g$.
The above-stated expressions, however, are applicable to generic hydrogenic
states.

\end{document}